\documentstyle[12pt]{article}
\begin{document}
\def\th{\theta}
\def\de{\Delta}
\def\cj{{\Im}}
\def\kb{k_\beta}
\def\beq{\begin{equation}}
\def\enq{\end{equation}}
\def\beqn{\begin{eqnarray}}
\def\eenq{\end{eqnarray}}
\def\pl{\parallel}
\baselineskip20pt

\begin{flushright}
MRI-PHY/96-18 
\end{flushright}

\begin{center}
{\large{\bf d-Wave Order Parameter in Bi2212 from a  
    Phenomenological Model of High T$_c$ Cuprates}}\\
\vspace{0.8cm} 
{\bf Biplab Chattopadhyay}\\
{Mehta Research Institute, 10 Kasturba Gandhi Marg, 
                                Allahabad 211002, INDIA} 
\end{center} 
\vspace{1cm}

\begin{abstract}

A phenomenological lattice model of high $T_c$ cuprates including 
order parameter phase fluctuations is considered within the BCS 
approximation, to interpret the experimental data from ARPES 
measurements on Bi2212 samples. A Kosterlitz-Thouless (KT) transition 
temperature $T_c^{KT}$ is estimated below the mean field transition 
$T_c^{MF}$, phase boundaries between competing order parameters of 
different symmetries are obtained and best model parameters, fitting 
the ARPES gap of $d_{x^2-y^2}$ symmetry, are determined. Variation 
of $T_c^{KT}$, as a function of the dopant concentration $\delta$, 
is in qualitative agreement with experiments. 
\end{abstract}
\vspace{1ex} 
\noindent{PACS numbers: 74.20.Mn, 74.20.Fg, 74.72.-h} 

\vspace{1ex} 
{\noindent Keywords: Helicity Modulus, Kosterlitz-Thouless 
                     Transition, Cuprate Superconductors,\\ 
~\indent~~~~~~~~~~Order Parameter Symmetry \hfill

\vfill
\noindent \rule{15cm}{0.1mm}\\
{\small email: biplab@mri.ernet.in\,;~~Fax: 0091-532-607991\,;
         ~~Phone: 0091-532-609097}

\newpage 

The symmetry of the superconducting order parameter (OP) in the 
high $T_c$ cuprate materials has been a widely debated issue in 
the last few years. In the recent past, several experiments done 
on the copper oxide materials [1-7], including those of phase 
sensitive experiments \cite{ex2}, interference measurements 
\cite{ex3} and c-axis Josephson tunneling \cite{ex4}, gave contrasting 
or inconclusive results regarding the OP symmetry. While a number 
of experiments \cite{ex1,ex2,ex5} noted signatures of $d_{x^2-y^2}$ 
OP symmetry, others argued in favour of the anisotropic $S$-wave 
or more exotic $S+\imath d$ wave OP symmetry \cite{ex3,ex4,ex6,ex7}. 
However, recently there has been considerable progress in the angle 
resolved photoemission spectroscopy (ARPES) measurements 
\cite{mrand,newexp}, and a consensus seems to be emerging about the 
OP symmetry \cite{review} in the high $T_c$ cuprates. 

ARPES can give quantitative estimate of the momentum dependence of 
the superconducting gap on the Fermi surface (FS) in terms of the 
spectral function representation of data and a detailed study of the 
FS is also possible \cite{mrand,newexp}. Due to its high angular as 
well as energy resolution \cite{ex6,newexp}, it can provide detailed 
and reliable knowledge about the nodes of the gap on the FS. Shen and 
coworkers, by ARPES measurements on Bi2212 compounds \cite{ex5}, found 
nodes of the superconducting gap on the FS along the 45 $^o$  
($\pi,\,~\pi$) direction, suggestive of a d-wave OP symmetry. Ding 
et. al., by similar measurements \cite{ex6} on high quality Bi2212 
single crystal, showed that the gap on the FS vanishes at two points 
(per quadrant) symmetrically displaced about the 45$^o$ direction, 
consistent with an anisotropic S-wave ($S_{xy}$) symmetry. However, 
reanalysis of their data revealed that the two node gap was an artifact 
of the superstructure producing ``ghost" bands \cite{norman}. In this 
light, the ARPES measurements on several Bi2212 samples were redone by 
Ding et. al. using dense sampling of the Brillouin zone (BZ) in the 
vicinity of the FS \cite{newexp}. The results are now consistent with 
a $|cos(k_x) - cos(k_y)|$ type gap function implying a $d_{x^2-y^2}$ 
OP symmetry. 

In order to interpret the two node gap data \cite{ex6}, a 
phenomenological BCS like lattice model in two dimension was introduced 
\cite{fenor} and mean field (MF) analysis of the instabilities in the 
spin singlet Cooper channel was done. The model is of interacting 
electrons on a square lattice, with an on-site repulsive ($V_0$) and 
attractive nearest neighbour ($V_1$) as well as next nearest neighbour 
($V_2$) interactions. The model Hamiltonian 
${\rm H} = {\rm H}_0 + {\rm H}_1$ is   

$$
 H_0 = \sum_{k,\sigma} (\epsilon_{\vec{k}} - \epsilon_F) 
      {c_{\vec{k},\sigma}^\dagger} {c_{\vec{k},\sigma}}\eqno(1a)\\ 
$$   
and 
$$  
  H_1  = V_0 \sum_{i} \hat{n}_{i,\uparrow} \hat{n}_{i,\downarrow} 
       + V_1 \sum_{i,\sigma,\sigma^\prime} \sum_{\delta_{nn}}
    \hat{n}_{i,\sigma} 
    \hat{n}_{i+\delta_{nn},\sigma^\prime}   
       + V_2 \sum_{i,\sigma,\sigma^\prime} \sum_{\delta_{nnn}}
    \hat{n}_{i,\sigma} 
       \hat{n}_{i+\delta_{nnn},\sigma^\prime}\eqno(1b)\\ 
$$  
Where, $c_{\vec{k},\sigma}^\dagger$ ($c_{\vec{k},\sigma}$) is the 
quasiparticle creation (annihilation) opeator of momentum $\vec{k}$ 
and spin $\sigma$, $\hat{n}$ is the number operator and $\delta_{nn}$, 
$\delta_{nnn}$ are nearest neighbour and next nearest neighbour 
lattice vectors. The band dispersion $\epsilon_{\vec{k}}$ was obtained 
by a six parameter tight binding fit to the normal state ARPES data 
on Bi2212 single crystal \cite{fenor} where the parameters are 
[$t_0,...,t_5$] = [0.131,\,-0.149,\,0.041,-0.013,\,-0.014,\,0.013] 
(in eV). Here $t_0$ is the orbital energy, $t_1$ nearest neighbour (nn), 
$t_2$ next nearest neighbour (nnn) etc. hopping matrix elements. 
Quasiparticle dispersion $\epsilon_{\vec{k}}$ incorporates flat 
dispersion around $(0,\,\pi)$ and $(\pi,\,0)$ points which results  
van Hove singularity (vHS) in the single particle density of states (DOS). 
\addtocounter{equation}{1} 

The mean field analysis of the model \cite{fenor} found strong 
instabilities for the $d_{x^2-y^2}$ and $S_{xy}$ OPs which can best 
exploit the large single particle DOS just below the FS. Ratio of the 
interaction parameters $V_1/V_2$ determines the relative stability of 
these two states. Further extension of the mean field work, to include 
strong fluctuations present in the quasi two dimensional cuprate 
materials, was carried out and results were reported in a previous 
communication \cite{biplab}. 

In this paper, we reanalyse the phenomenological model, including 
order parameter phase fluctuations, in view of the conclusive finding 
of a $d_{x^2-y^2}$ OP symmetry by the ARPES measurements on Bi2212 
\cite{ex5,newexp}. We first calculate a Kosterlitz-Thouless (KT) 
transition temperature $T_c^{KT}$, using the helicity modulus or 
superfluid phase stiffness ($\rho_s$) expression of the present model 
together with the KT relation $\rho_s(T_c^{KT}) = {2\over\pi}k_BT_c^{KT}$, 
within each of the irreducible representation B$_1$ ($d_{x^2-y^2}$), 
B$_2$ ($d_{xy}$) and A$_1$ ($S,\,S^\star~{\rm and}~S_{xy}$). We then 
find out the phase diagrams showing regions of relative stability of 
B$_1$, B$_2$ and A$_1$ states in the interaction parameter planes and 
determine the best model parameters corresponding to the $d_{x^2-y^2}$ 
gap with $T_c^{KT} \sim 100\,K$. We also study the momentum dependence 
of the gap function on the FS and dopant concentration ($\delta$) 
dependence of $T_c^{KT}$. Our main results are summarized below:   

\begin{itemize} 
\item[(a)] Best model parameters at the optimal doping level 
           $\delta = 0.17$, with a $T_c^{KT} \sim 100\,K$ and for the 
           stable B$_1$ ($d_{x^2-y^2}$) state, are $V_0 \ge 400\,meV$, 
           $V_1 \approx -48\,meV$ and $V_2 \sim -85\, meV$. 
\item[(b)] Zero temperature $d_{x^2-y^2}$ gap on the FS, with 
           $V_1 = -48\,meV$, match well the ARPES data \cite{newexp} in 
           the vicinity of 45$^o$ direction, but deviates from it as one 
           moves beyond about 7 degrees on either sides.   
\item[(c)] With varying dopant concentration ($\delta$), the gap magnitude 
           at 0$^o$ angular direction changes considerably, although the node 
           position is same for all $\delta$ by virtue of the momentum 
           dependent part of the $d_{x^2-y^2}$ gap function. 
\item[(d)] $T_c^{KT}$ for $B_1$ ($d_{x^2-y^2}$) state as a function of 
           $\delta$ shows correct qualitative behaviour as in the high 
           $T_c$ cuprate materials. 
\end{itemize} 

Within the standard BCS approximation, the Hamiltonian of Eq.(1) yields 
the gap equation 
\beq
\Delta_{\vec{k}}={1\over N} \sum_{\vec{k}^\prime} V(\vec{k}-\vec{k}^\prime) 
          {{\Delta_{\vec{k}^\prime}}\over {2E_{\vec{k}^\prime}}} 
          \tanh \left({{\beta E_{\vec{k}^\prime}}\over 2}\right) 
\enq 
where the quasiparticle energy is $E_{\vec{k}} = 
\sqrt{(\epsilon_{\vec{k}}-\epsilon_F)^2 + |\Delta_{\vec{k}}|^2}$ and  
$\Delta_{\vec{k}}$ is the BCS gap function. 
The pairing interaction, 
$V(q) = V_0 + 4 V_1(\cos q_x + \cos q_y) + 8 V_2 \cos q_x \cos q_y$,  
is separable as $V(\vec{k} - \vec{k}^\prime) 
= \sum_{i=0}^4 \tilde{V}_i \eta_i(\vec{k}) \eta_i(\vec{k}^\prime) $. 
An expansion of the order parameter 
$\Delta_{\vec{k}} = \sum_{i=0}^4 \eta_i(\vec{k}) \Delta_i$ gives the 
linearized gap equation 

\begin{equation}  
\Delta_i = -{\tilde{V}_i\over 2N} \sum_{\vec{k}} 
     {\eta_i(\vec{k})\over E_{\vec{k}}}\tanh ({\beta E_{\vec{k}}\over 2}) 
     \sum_{j\in\cal{R}}\Delta_j \eta_j(\vec{k})  
\end{equation}
where $\eta_0(\vec{k}) = 1$, 
$\eta_1(\vec{k}) = \frac{1}{2}(\cos k_x + \cos k_y)$, 
$\eta_2(\vec{k}) = \frac{1}{2}(\cos k_x - \cos k_y)$, 
$\eta_3(\vec{k}) = \cos k_x\,\cos k_y$, 
$\eta_4(\vec{k}) = \sin k_x\,\sin k_y$ are the basis functions 
corresponding to [$S,\, S^\star,\, d_{x^2-y^2},\,$ $S_{xy},\, d_{xy}$] 
symmetries, and 
$(\tilde{V}_0...\tilde{V}_4) = (V_0, 8V_1, 8V_1, 8V_2$, $8V_2)$. 
We suppress writing the terms corresponding to triplet pairing and 
ignore them in our analysis. Here ${\cal{R}} \equiv (A_1, B_1, B_2)$ 
are different irreducible representations of the $C_{4v}$ group and 
the gap equation factorizes to independent $\Delta_2$ (B$_1$ representation), 
$\Delta_4$ (B$_2$ representations) and three coupled linear equations  
$\Delta_0,\,\Delta_1,\,\Delta_3$ (A$_1$ representation). 
Since $\Delta_3$ is the predominant component within the A$_1$ 
representation, it is identified as a state of $S_{xy}$ OP symmetry. 

To calculate the helicity modulus $\rho_s$ (and hence $T_c^{KT}$ 
thereafter), a transverse vector potential with the gauge $A_y = 0$ 
is considered. This introduces an extra phase which the carriers 
acquire while moving between the lattice sites. Hence the hopping 
matrix elements in $H_0$ (Eq.(1a)) should be changed through Peierls 
substitution $t_{ij}\rightarrow t_{ij}\exp[{{{ie}\over{\hbar c}}
\int_{{\vec R}_j}^{{\vec R}_i}{\vec{A}}\cdot {\vec dl}}]$.
We work here with the units $\hbar = c = 1$, but explicitly write 
them whenever necessary. 

The electron current operator ${\hat{j}}_x(\vec{R}_i)$ consists of the 
usual paramagnetic and diamagnetic terms \cite{schrie}. To linear order 
in $A_x$, ${\hat{j}}_x(\vec{R}_i)$ is obtained by differentiating $H_0$ 
with respect to $A_x(\vec{R}_i)$  
$$
{\hat{j}}_x(\vec{R}_i) = -c {{\partial H_0}\over{\partial A_x(\vec{R}_i)}}  
     = {\hat{j}}_x^{para}(\vec{R}_i) + {\hat{j}}_x^{dia}(\vec{R}_i)\eqno(4)
$$ 
In Eq.(4), the paramagnetic term does not involve $A_x$ and is the 
electron velocity operator. The diamagnetic term is linear in $A_x$ and
stems from the Meissner screening of the condensate. Average value of 
the diamagnetic current density is obtained as   
$$ 
j_x^{dia}(\vec{q}) = -{{e^2}\over{\hbar^2 c}} {1\over N} 
   \sum_{\vec{k},\sigma} \left\langle 
   c_{\vec{k},\sigma}^\dagger  c_{\vec{k},\sigma} \right\rangle 
   {{\partial^2\epsilon_{\vec{k}}}\over{\partial k_x^2}} 
   A_x(\vec{q})\eqno(5)                                          
$$ 
where the $\langle\;\,\,\rangle$ represents an average in the mean 
field superconducting state. In a London like relation 
$j_x(\vec{q})\propto -\rho_s A_x({\vec{q}})$, the diamagnetic 
contribution to the phase stiffness ($\rho_s^{dia}$) is proportional to 
${1\over N}\sum_{\vec{k},\sigma}\left\langle 
c_{\vec{k},\sigma}^\dagger  c_{\vec{k},\sigma} \right\rangle 
{{\partial^2\epsilon_{\vec{k}}}\over{\partial k_x^2}} $,  the mean 
electronic kinetic energy along the x-direction \cite{scala}. 
The average in the lattice model turns out to be 
$\left\langle c_{\vec{k},\sigma}^\dagger  c_{\vec{k},\sigma} \right\rangle 
= {1\over 2} \left[ 1- {{\epsilon_{\vec{k}}-\epsilon_F}\over{E_{\vec{k}}}} 
\tanh \left({\beta E_{\vec{k}}\over 2}\right) \right]$, unlike the 
continuum case where $\rho_s^{dia}\propto -\sum_{\vec{k},\sigma} 
\left({{\partial\epsilon_{\vec{k}}}\over{\partial k_x}}\right)^2 {{\partial 
f(\epsilon _{\vec{k}}-\epsilon_{F})}\over{\partial \epsilon_{\vec{k}}}}$ 
($f$ is the Fermi function).
\addtocounter{equation}{2} 

Contribution of the paramagnetic part is evaluated using linear response 
theory. In the long wavelength limit, the paramagnetic current is found 
to be 
$$ 
j_x^{para}(\vec{q}) = - {1\over c} \left[\lim_{\vec{q}\to0} \lim_{\omega\to0}
     K^{xx}(\vec{q},\omega)\right]A_x(\vec{q})\eqno(6) 
$$ 
where 
\addtocounter{equation}{1} 
$K^{xx}(\vec{q},\omega) = -\imath \int dt\, \th(t)\,
    e^{\imath\omega t} \left\langle\left[ 
    j_x^{para}(\vec{q},t),\,j_x^{para}(-\vec{q},0)
    \right]\right\rangle $. 
The correlation function in Eq.(6) is evaluated to be 
$K^{xx}({{\vec{q}}\to0},{\omega\to0}) 
= {{e^2}\over{\hbar^2}} {1\over N} \sum_{\vec{k}} 
 \left({{\partial\epsilon_{\vec{k}}}\over{\partial k_x}}\right)^2 
 {{\partial f(E_{\vec{k}})}\over{\partial E_{\vec{k}}}}$. 
Taking the contributions from diamagnetic and paramagnetic parts, from 
Eqs.(5) and (6), we obtain the expression for superfluid phase stiffness 
\beq 
\rho_s(T) = {1\over{2N}} \sum_{\vec{k}} \left[ 
  \left({{\partial\epsilon_{\vec{k}}}\over{\partial k_x}}\right)^2 
  {{\partial f(E_{\vec{k}})}\over{\partial E_{\vec{k}}}}  + {1\over 2} 
  {{\partial^2\epsilon_{\vec{k}}}\over{\partial k_x^2}} 
  \left\{ 1- {{\epsilon_{\vec{k}}-\epsilon_F}\over{E_{\vec{k}}}} 
  \tanh \left({\beta E_{\vec{k}}\over 2}\right) \right\} \right]. 
\enq 
It should me mentioned here that, we work in a transverse gauge and 
vertex corrections required to get a gauge invariant current \cite{schrie} 
have not been included. 

Above expression for $\rho_s$ involves linearized BCS gaps. In the 
inset of Fig.1 we plot the superconducting gaps for different order 
parameters corresponding to $B_1$, $B_2$ and $A_1$ states, as a 
function of temperature. The point where gaps become nonzero mark 
the mean field transition $T_c^{MF}$ for a state. These gaps, as shown  
in the inset of Fig.1, are used to calculate $\rho_s(T)$ for different 
states. The KT transition temperature for each state is found by 
comparing the $\rho_s(T)$ curve for each state, with the KT relation 
$\rho_s(T_c^{KT}) = {2\over\pi}T_c^{KT}$. In Fig.1 the intersecting 
point of a $\rho_s$ curve with the KT straight line (emerging from the 
origin) gives $T_c^{KT} < T_c^{MF}$. Similar technique was applied 
previously by Danteneer and coworkers \cite{denten} for the two 
dimensional attractive Hubbard model and correct behaviour of $T_c^{KT}$, 
including its inverse coupling dependence in the strong coupling limit, 
was found. We too find a similar strong coupling dependence of $T_c^{KT}$ 
in the present model. Thus, order parameter phase fluctuation degrades 
the mean field transition temperature. However, $T_c^{KT}$ in our 
case is an upper bound of the actual KT transition temperature, 
that could be calculated only by considering superconducting gap 
renormalization due to the presence of vortex like fluctuations. 

Next, we consider the phase boundaries, calculated by comparing the
$T_c^{KT}$ values of the competing states. In Fig.2, we plot phase 
boundaries separating $d_{x^2-y^2}$ and $A_1$ states in the 
(-$V_1$, $-V_2$) plane, for various values of $V_0$. It clearly shows 
an widening of the the $d_{x^2-y^2}$ stable region with increasing on 
site repulsion for small values of $-V_1$. Rate of this widening is 
faster for small $V_0$. For a fixed $-V_1$, $d_{x^2-y^2}$ solution is 
preferred  over $A_1$ upto a maximum $V_2$. As for example, with 
$V_0=100\,meV$ and $V_1 = -48\,meV$, $d_{x^2-y^2}$ solution is stable 
for $-V_2\leq 80\,meV$ \cite{comt1}. A comparison of the KT phase diagrams 
with those from mean field calculations, can be found in Ref.[13]. 

To determine the optimal model parameters, one must also consider the 
phase boundary between $A_1$ and $d_{xy}$ states and find out the 
region where $A_1$ solution is not stable. In Fig.3, we plot such phase 
boundaries in the ($-V_2,\,V_0$) plane for different values of $V_1$. 
To get the KT transition temperature $T_c^{KT} \sim 100\,K$ for 
$d_{x^2-y^2}$ OP, one must have $V_1 = -48\,meV$. In Fig.3, if we set 
$V_1 = -48\,meV$ and $V_2 = -85\,meV$, then to rule out the stability 
of $A_1$ state, one must have $V_0\ge 400\,meV$. Thus, we fix the 
optimal model parameters to be $V_1 = -48\,meV$, $V_2 = -85\,meV$ 
and $V_0 \ge 400\,meV$. 

With these optimal parameters, we study the momentum dependence of 
the zero temperature $d_{x^2-y^2}$ gap on the FS at $\delta = 0.17$. 
A plot of $|\Delta_{\vec{k}}(T=0)| = |\Delta_2(T=0) \times {1\over 2} 
[\cos(k_x) - \cos(k_y)]|$ is presented in Fig.4 for two different 
values of $V_1$. Solid circles are ARPES data \cite{newexp} of 
$T_c \sim 87\,K$ Bi2212 single crystal sample in the Y quadrant 
\cite{comt2}. We find that, our $V_1 = -48 \,meV$ curve matches well 
the experimental data till about 7$^o$ away from node position 45$^o$ 
on either sides, and falls below ARPES data beyond this. A much better 
match could be obtained with $V_1 = -58\,meV$. But, this makes 
$T_c^{KT} \sim 135\,K$ which is well above the sample $T_c\sim 87\,K$. 
One probable argument in favour of this could be that, our $T_c^{KT}$ 
is an upper bound of the KT transition temperature. In actual case, 
gap renormalization due to the vortex fluctuations might reduce 
$T_c^{KT}$. If one assumes a 25$\%$ reduction, a new set of best model 
parameters can be obtained as $V_1 = -58\,meV$, $V_2 = -90\,meV$ and 
$V_0 \ge 400\,meV$. In the inset of Fig.4, we plot the $\delta$ 
dependence of the zero temperature gap which provides a feel for the 
$d_{x^2-y^2}$ gap magnitude at the FS angle $\phi = 0$ degree. 

Variation of $T_c^{KT}$ for $d_{x^2-y^2}$ with dopant concentration 
$\delta$ is presented in Fig.5. This shows a correct qualitative bell 
shaped behaviour as in the high $T_c$ experiments. As a comparison, 
we also include $T_c^{KT}$ for $A_1$ state for the optimal parameters, 
which is below the $d_{x^2-y^2}$ $T_c^{KT}$ curve for all $\delta$. The 
$T_c^{KT}$ curve here (and also $\Delta_2$ vs $\delta$ curve in the 
inset of Fig.4) peaks at around $\delta = 0.25$, that is consistent 
with the peak in the pairing density of states around 
$\epsilon_{vHS} \approx -20\,meV$ which  corresponds to 
$\delta \simeq 0.25$. 

To summarize, we studied a phenomenological BCS model of high $T_c$  
cuprate superconductors including fluctuations, calculated phase 
boundaries separating order parameters of different symmetries within 
the model and determined optimal model parameters to fit the 
$d_{x^2-y^2}$ superconducting gap as in the experiment. We also studied 
the dopant concentration dependence of the transition temperature which 
is in qualitative agreement with the experiments. 

Its a pleasure to acknowledge Dattu Gaitonde for useful discussions 
and suggestations.

\newpage 

\noindent {\bf FIGURE CAPTIONS}

\noindent Fig.1.  Helicity modulus $\rho_s$ is plotted as a function of 
 temperature for different irreducible representations, at optimal doping 
 $\delta=0.17$ and with best model parameters 
 $V_0=400\,meV,\,\, V_1=-48\,meV$ and $V_2=-85\,meV$. 
 The solid straight line, originating from (0,0) point, is the 
 $\rho_s(T_c^{KT}) = {2\over\pi}T_c^{KT}$ line. [Inset: Mean field order 
 parameters of different symmetries corresponding to $A_1$, $B_1$ and 
 $B_2$ representations.] 

\noindent Fig.2.  Phase boundaries, separating the regions of stability 
 of the $A_1$ and $d_{x^2-y^2}$ states, in the ($-V_1,-V_2$) plane, for 
 $\delta = 0.17$ and various $V_0$ values as shown in the figure. 
 [Inset: Phase diagram for $V_0 = 0.4\,eV$ is plotted on larger scales.] 

\noindent Fig.3.  Phase boundaries indicating the regions of stability of 
 the $A_1$ and $d_{xy}$ states in the ($-V_2,V_0$) plane at $\delta=0.17$ 
 for various $V_1$ values shown. Inset shows the phase diagram for 
 $V_1=-48\,meV$ on an expanded scale. 

\noindent Fig.4.  Momentum dependence of the gap function 
 $|\Delta_{\bf k}(T=0)|$ on the Fermi surface is plotted against the  
 Fermi surface angle $\phi$, for $\delta=0.17$ and for $V_1$ given in the 
 figure. The angle $\phi$ is measured with respect to the line joining 
 $(\pi,\pi)$ and $(0,\pi)$ points. Solid circles are 
 ARPES data from Ref.[9]. [Inset: Zero temperature $d_{x^2-y^2}$ gap 
 as a function of $\delta$.] 

\noindent Fig.5. Kosterlitz-Thouless transition temperature $T_c^{KT}$ 
 for $d_{x^2-y^2}$ and $A_1$ states as a function of dopant concentration 
 $\delta$, for optimal model parameters. 
     

\begin{thebibliography}{999}
\bibitem{comt1}{Below some values of $-V_1$ and $-V_2$, KT transition 
  temperature $T_c^{KT} < 1\,K$ and its not possible to determine the 
  phase boundaries beyond these parameter values.}
\bibitem{comt2} {In Y quadrant, the main CuO FS is widely separated from 
  the two ghost FS and gap data are more reliable. Hence, Y quadrant data 
  from Ref.[9] is chosen for comparison.}
\end{thebibliography}
\end{document}